\documentclass[usegraphicx,usenatbib,letterpaper]{emulateapj}

\usepackage{ulem}
\usepackage{color}
\usepackage{graphics,graphicx}
\usepackage{times} 
\usepackage{amssymb}
\usepackage{amsmath}
\usepackage{natbib}
\usepackage{url}
\usepackage{multirow}
\usepackage{ctable}
\usepackage{threeparttable}
\newif\ifAMStwofonts
\AMStwofontstrue

\def\xmm{{\it XMM-Newton}}

\def\epicpn{{EPIC-pn}}
\def\epicmos1{{EPIC-MOS1}}
\def\epicmos2{{EPIC-MOS2}}
\def\epicmos{{EPIC-MOS}}

\def\nustar{{\it NuSTAR}}

\def\kmps{\hbox{$\rm\thinspace km~s^{-1}$}}

\def\H0{{\rm ~km~s^{-1}~Mpc^{-1}}}

\def\kev{\hbox{\rm keV}}

\def\photpkevpcmsqps{\hbox{$\rm\thinspace ct~keV^{-1}~cm^{-2}~s^{-1}$}}

\def\ergps{\hbox{erg~s$^{-1}$}}
\def\ergcmps{\hbox{\rm erg~cm~s$^{-1}$}}

\def\msun{\hbox{$\rm M_{\odot}$}}

\def\chisq{{$\chi^{2}$}}

\def\xspec{\hbox{\small XSPEC}}
\def\xspecv{\hbox{\small XSPEC}\, v12.8.2}

\def\nustardas{\rm {\small NUSTARDAS}}

\def\addascaspec{\hbox{\rm{\small ADDASCASPEC~\/}}}

\def\epchain{\hbox{\rm{\small EPCHAIN}}}
\def\emchain{\hbox{\rm{\small EMCHAIN}}}

\def\rmfgen{\hbox{\rm{\small RMFGEN}}}
\def\arfgen{\hbox{\rm{\small ARFGEN}}}

\def\addascaspec{\hbox{\rm{\small ADDASCASPEC}}}
\def\nupipeline{\rm{\small NUPIPELINE}}
\def\nuproducts{\rm{\small NUPRODUCTS}}

\def\xstar{\hbox{\rm{\small XSTAR}}}
\def\grid25{\hbox{\rm{\small GRID25}}}

\def\tbabs{\rm{\small TBABS}}

\def\fexxv{\hbox{\rm Fe\,{\small XXV}}}
\def\fexxvi{\hbox{\rm Fe\,{\small XXVI}}}

\def\ka{K\,$\alpha$}

\def\eg{{\it e.g.}}

\def\ie{{\it i.e.~\/}}

\def\la{\mathrel{\hbox{\rlap{\hbox{\lower4pt\hbox{$\sim$}}}{\raise2pt\hbox{$<$}}}}}
\def\ga{\mathrel{\hbox{\rlap{\hbox{\lower4pt\hbox{$\sim$}}}{\raise2pt\hbox{$>$}}}}}

\def\d25{D$_{25}$}

\def\.25{0.25 keV\thinspace}

\def\ngc{\rm NGC\,1313 X-1}
\def\elow{\rm 6.6}
\def\ehigh{\rm 9.6}
\def\nsims{\rm 10,000}
\def\ngtr{\rm 31}
\def\ngtrexcl{\rm 12}

\shorttitle{The UFO in NGC\,1313 X-1: Fe K}
\shortauthors{D.~J. Walton et al.}

\begin{document}

\title{An Iron K Component to the Ultrafast Outflow in NGC\,1313 X-1}

\author{D. J. Walton\altaffilmark{1,2},
M. J. Middleton\altaffilmark{3},
C. Pinto\altaffilmark{3},
A. C. Fabian\altaffilmark{3},
M. Bachetti\altaffilmark{4},
D. Barret\altaffilmark{5,6},
M. Brightman\altaffilmark{2},
F. Fuerst\altaffilmark{2}, \\
F. A. Harrison\altaffilmark{2},
J. M. Miller\altaffilmark{7},
D. Stern\altaffilmark{1},
}
\affil{
$^{1}$ Jet Propulsion Laboratory, California Institute of Technology, Pasadena, CA 91109, USA \\
$^{2}$ Space Radiation Laboratory, California Institute of Technology, Pasadena, CA 91125, USA \\
$^{3}$ Institute of Astronomy, University of Cambridge, Madingley Road, Cambridge CB3 0HA, UK \\
$^{4}$ INAF/Osservatorio Astronomico di Cagliari, via della Scienza 5, I-09047 Selargius (CA), Italy \\
$^{5}$ Universite de Toulouse; UPS-OMP; IRAP; Toulouse, France \\
$^{6}$ CNRS; IRAP; 9 Av. colonel Roche, BP 44346, F-31028 Toulouse cedex 4, France \\
$^{7}$ Department of Astronomy, University of Michigan, 1085 S. University Ave., Ann Arbor, MI, 49109-1107, USA \\
}

\begin{abstract}
We present the detection of an absorpton feature at $E=8.77^{+0.05}_{-0.06}$\,\kev\
in the combined X-ray spectrum of the ultraluminous X-ray source \ngc\ observed
with \xmm\ and \nustar, significant at the 3$\sigma$ level. If associated with
blueshifted ionized iron, the implied outflow velocity is $\sim$0.2$c$ for \fexxvi, or
$\sim$0.25$c$ for \fexxv. These velocities are similar to the ultrafast outflow seen
in absorption recently discovered in this source at lower energies by \xmm, and
we therefore conclude that this is an iron component to the same outflow.
Photoionization modeling marginally prefers the \fexxv\ solution, but in either case
the outflow properties appear to be extreme, potentially supporting a
super-Eddington hypothesis for \ngc.
\end{abstract}

\begin{keywords}
{Black hole physics -- X-rays: binaries -- X-rays: individual (\ngc)}
\end{keywords}

\section{Introduction}

Ultraluminous X-ray sources (ULXs) are variable, off-nuclear point sources in
nearby galaxies with X-ray luminosities $L_{\rm{X}}\geq10^{39}$\,\ergps\
(\citealt{Swartz04, WaltonULXcat}). The brighter members of this population
have luminosities that significantly (factors of 10 or more) exceed the Eddington
limit for the $\sim$10\,\msun\ stellar-remnant black holes observed in accreting
Galactic black hole binaries (\citealt{Casares14rev}). Multi-wavelength
observations have largely ruled-out strong anisotropic emission as a means of
skewing luminosity estimates (\citealt{Moon11}, although moderate collimation is
still permitted). ULXs must therefore either host large black holes, potentially either
the long-postulated yet elusive `intermediate mass' black holes
($M_{\rm{BH}}\sim10^{2-5}$\,\msun; \citealt{Miller04}) or the massive stellar
remnants ($M_{\rm{BH}}\sim30-100$\,\msun; \citealt{Zampieri09}) recently
confirmed by LIGO (\citealt{Abbott16gw}), or represent an exotic, highly
super-Eddington accretion phase (\citealt{Poutanen07}). In either case, they
hold clues to the processes governing the formation and evolution of
supermassive black holes in the early Universe (\citealt{Kormendy13rev}).

Since launch, the \nustar\ mission (\citealt{NUSTAR}) has undertaken a substantial
program observing a sample of extreme ULXs, revealing the high-energy
($E>10$\,keV) behavior of these enigmatic sources for the first time. As one of the
few sources within $\sim$5\,Mpc to persistently radiate at
$L_{\rm{X}}\sim10^{40}$\,\ergps\ (\citealt{Miller13ulx}), \ngc\ ($D\sim4$\,Mpc)
was an important part of this program, observed in coordination with \xmm\
(\citealt{XMM}) to provide broadband ($\sim$0.3--30\,\kev) spectral coverage. These
observations revealed broadband spectra inconsistent with standard modes of
sub-Eddington accretion (\citealt{Bachetti13, Miller14}; similar to other ULXs
observed by \nustar\ to date, \citealt{Walton14hoIX, Walton15, Walton15hoII,
Rana15, Mukherjee15}), supporting the idea that these sources are exhibiting a
super-Eddington phase of accretion. Indeed, we now know at least one of these
sources is a highly super-Eddington neutron star (\citealt{Bachetti14nat}).

A prediction of all super-Eddington accretion models is that powerful
winds should be launched (\citealt{Poutanen07, King09, Dotan11, Takeuchi13}).
Robust detection of any such winds from ULXs has, however, proven challenging
(\citealt{Walton12ulxFeK, Walton13hoIXfeK}). For \ngc, \cite{Middleton15soft} report
low-energy ($\sim$1\,keV) blended atomic features that are consistent with being
absorption from an ionised outflow, but the low-resolution CCD spectra considered
prevented a conclusive identification as such. However, in a key breakthrough, a
recent follow-up analysis by \cite{Pinto16nat} utilizing the high-resolution reflection
grating spectrometer (RGS) aboard \xmm\ was able to resolve this low-energy
spectral structure into several discrete emission and absorption features, and found
that \ngc\ does indeed exhibit an extreme ionised outflow, potentially consisting of
multiple velocity components spanning $\sim$0.2--0.25$c$.

Here, by considering the high-energy \xmm\ and \nustar\ data available for \ngc,
we report on a detection of an ionized iron \ka\ component to the ultrafast
outflow (UFO) discovered by \cite{Pinto16nat}.

\section{Observations and Data Reduction}
\label{sec_red}

\ngc\ has frequently been observed in the X-ray band and is known to be a
variable source, exhibiting high- and low-states with significantly different spectra
(\citealt{Feng06, Pintore12, Middleton15}, Bachetti et al. in preparation). The
majority of the X-ray observations in the archive cover the low-flux state, so we
focus on these data in order to maximise the integrated signal-to-noise (S/N) in
the iron K bandpass while considering only observations with similar spectra.
\xmm, in particular, has frequently observed NGC\,1313, but many of these
observations are short and have X-1 placed off-axis. The \epicpn\ detector aboard
\xmm\ is known to have increased background emission from copper lines at
$\sim$8\,\kev\ away from the optical axis (\citealt{XMMblanksky}), which fall in the
energy range of interest for any iron line searches given the velocities reported by
\cite{Pinto16nat}. We therefore also limit our analysis to \xmm\ observations where
\ngc\ was the primary target. In total, we consider the three full-orbit \xmm\
observations (OBSIDs 0405090101, 0693850501, 0693851201; note that these
are the same \xmm\ observations considered by \citealt{Pinto16nat}), and the two
long \nustar\ observations presented in \citet[OBSIDs 30002035002 and
30002035004]{Bachetti13}.

The data from these observations are reduced individually, and then combined
into a set of average spectra using \addascaspec. In all cases, source products
are extracted from a circular region of radius 40$''$ since there is another X-ray
source $\sim$55$''$ to the south (\citealt{Bachetti13}). This aperture gave the
best balance between maximising the S/N for X-1 and minimising contamination
from this other source. Background is always estimated from significantly larger
regions of blank sky on the same detector as X-1 to ensure it is well sampled.
Finally, each of the average spectra are rebinned to a minimum of 25 counts per
bin to allow the use of \chisq\ minimization during spectral fitting. The following
sections provide further technical details regarding our reduction of these data.

\subsection{NuSTAR}

The \nustar\ data were reduced using the standard pipeline, \nupipeline, part of the
\nustar\ Data Analysis Software (v1.4.1). \nustar\ caldb v20150316 is
used throughout. The unfiltered event files were cleaned with the standard depth
correction, significantly reducing the internal background, and passages through
the South Atlantic Anomaly were removed. Source spectra and instrumental
responses were produced for both of the focal plane modules (FPMA/B) using
\nuproducts. In addition to the standard `science' data, we also extract the
`spacecraft science' data following \cite{Walton16cyg}, which in this case provides
$\sim$35\% of the total 360\,ks (per FPM) good exposure.

\subsection{XMM-Newton}

The \xmm\ data were reduced with the \xmm\ Science Analysis System (v14.0.0),
following the standard prescription.\footnote{http://xmm.esac.esa.int/} Raw data files
were cleaned using \epchain\ for \epicpn\ (\citealt{XMM_PN}), and \emchain\ for the
\epicmos\ detectors (\citealt{XMM_MOS}). Only single--double 
({\small{PATTERN}}$\leq$4) and single--quadruple ({\small{PATTERN}}$\leq$12)
events were considered for \epicpn\ and \epicmos, respectively. Periods of high
background were excluded, as were events close to edge/bad pixels
({\small{FLAG}}=0). Instrumental response files were generated with \rmfgen\ and
\arfgen, and the data from the two \epicmos\ detectors were combined together. The
total good exposure is 258\,ks for \epicpn, and 331\,ks for each \epicmos\ unit.

\begin{figure}
\hspace*{-0.35cm}
\epsscale{1.16}
\plotone{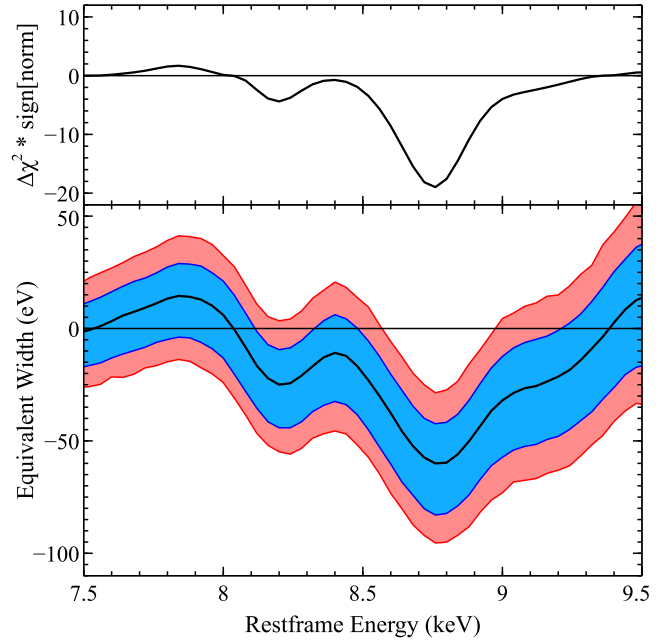}
\caption{A zoom in on the line-search results for the absorption feature detected in
our analysis. Top: the $\Delta$\chisq\ improvement obtained with the addition of a
narrow Gaussian line, as a function of line energy, for \ngc. Positive values indicate
the best fit line is in emission, and negative values indicate absorption. The feature
at $\sim$8.8\,keV gives an improvement of $\Delta\chi^{2}=19.5$. Bottom: 90\%
(\textit{blue}) and 99\% (\textit{red}) confidence contours for the equivalent width of
the narrow line included. We find this feature at $\sim$8.8\,keV to have
$\mathrm{EW}=61\pm24$\,eV. No other siginificant features are detected.}
\vspace{0.2cm}
\label{fig_search}
\end{figure}

\section{Spectral Analysis}
\label{sec_spec}

Our goal in this work is to search for any ionized iron \ka\ absorption features in
the X-ray spectrum of \ngc\ that might be associated with the UFO
discoverd by \cite{Pinto16nat}. We use \xspecv\ (\citealt{xspec}) for our spectral
analysis, and unless stated otherwise parameter uncertainties are quoted at 90\%
confidence for one parameter of interest. Our data selection is such that we can
apply a common model to all the datasets, accounting for differences in the
average \xmm\ and \nustar\ fluxes and residual cross-calibration uncertainties
between the detectors by allowing multiplicative constants to float between them,
fixing \epicpn\ to unity. The \epicpn\ and \epicmos\ detectors agree to within 5\%,
as do FPMB and FPMB, while the average \xmm\ and \nustar\ fluxes resulting
from our data selection only differ by $\sim$10\%, which is broadly similar to the
absolute cross-calibration differences seen between the two missions
(\citealt{NUSTARcal}). 

We begin by constructing a simple model for the continuum. In this work, we
focus on the 3--20\,keV bandpass, providing sufficient coverage to accurately model
the continuum local to the iron band with a simple model, while remaining
independent of the low-energy X-ray band in which the outflow was initially
discovered. The spectral curvature seen from \ngc\ over this energy range is well
established (\citealt{Stobbart06, Gladstone09, Bachetti13}), so we model the
continuum as a cutoff powerlaw. We also include neutral absorption, both from our
Galaxy ($N_{\rm{H,Gal}}=4.1\times10^{20}$\,cm$^{-2}$; \citealt{NH}) and intrinsic to
NGC\,1313 ($z=0.00157$). However, given the limited bandpass considered, we are
not particularly sensitive to the level of absorption seen towards \ngc, so we fix the
intrinsic column to $N_{\rm{H,int}}=2.7\times10^{21}$\,cm$^{-2}$ (\citealt{Miller13ulx}).
These neutral absorption components are modeled with \tbabs, adopting the
abundance set of \cite{tbabs} and cross-sections of \cite{Verner96}. This provides an
excellent fit to the 3--20\,\kev\ emission, with \chisq/degrees of freedom $=1634/1627$.
The photon index and high-energy cutoff obtained are $\Gamma=0.96\pm0.07$ and
$E_{\rm{cut}}=5.5^{+0.4}_{-0.3}$\,\kev, and the model normalisation is
$(4.0\pm0.2)\times10^{-4}$\,\photpkevpcmsqps\ (at 1\,keV).

To search for atomic features, we follow a similar approach to \cite{Walton12ulxFeK,
Walton13hoIXfeK}. We refer the reader to those works for a detailed description, but
in brief, we include a narrow (intrinsic width of $\sigma = 10$\,eV) Gaussian, and vary
its energy across the energy range of interest in steps of 40\,eV (oversampling the
\xmm\ energy resolution by a factor $\sim$4--5). The Gaussian normalisation can be
either positive or negative. For each line energy, we record the $\Delta\chi^{2}$
improvement in fit resulting from the inclusion of the Gaussian line, as well as the best
fit equivalent width ($EW$) and its 90 and 99\% confidence limits. These are
calculated with the {\small EQWIDTH} command in \xspec, using 10,000 parameter
simulations based on the best fit model parameters and their uncertainties. To be
conservative, we vary the Gaussian line energy between \elow\ and \ehigh\,\kev,
corresponding to a wide range of outflow velocities extending up to $>$0.25$c$
for \fexxvi.

\begin{figure}
\hspace*{-0.35cm}
\epsscale{1.16}
\plotone{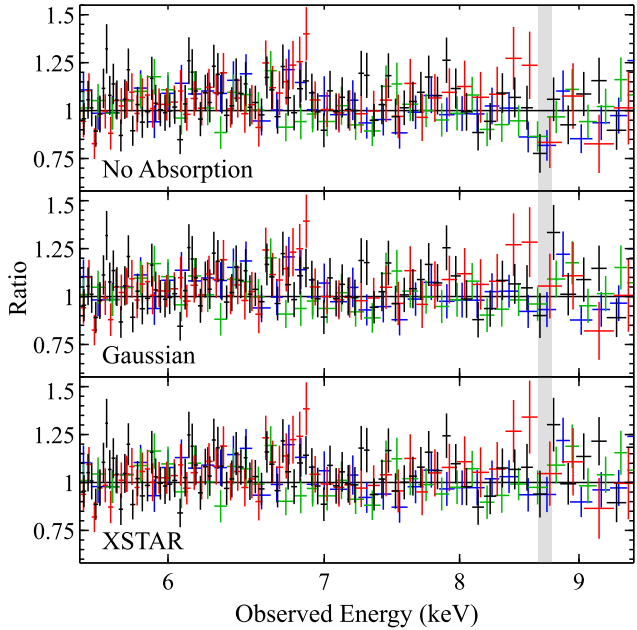}
\caption{
Data/model ratio plots for our basic continuum model (\textit{top panel}), the model
including a Gaussian absorption line (\textit{middle panel}) and the model including
a physical photoionised absorber (\xstar; \textit{bottom panel}). Data from \epicpn,
\epicmos, FPMA and FPMB are shown in black, red, green and blue, respectively.
The feature at $\sim$8.8\,keV is indicated with the grey shaded region; all of the
detectors used in this work show low residuals at the same energy, which are
improved in the models including a Gaussian and an \xstar\ absorption component.
}
\vspace{0.2cm}
\label{fig_ratio}
\end{figure}

The results are shown in Figure \ref{fig_search}. We find the addition of a Gaussian
absorption line at $\sim$8.8\,\kev\ provides a noteable improvement to the fit. Allowing
the line parameters to vary freely, we find a line energy of
$E=8.77^{+0.05}_{-0.06}$\,\kev, an equivalent width of $\mathrm{EW}=-61\pm24$\,eV
(comparable to the strongest iron absorption seen from a black hole binary to date;
\citealt{King12}). This gives an improvement to the fit of $\Delta\chi^{2}=19.5$ for
three extra free parameters. The line is consistent with being unresolved at the
resolution of the \xmm\ and \nustar\ detectors ($\sigma<0.24$\,\kev). Assuming an
association with iron, the extreme energy of this feature makes an association with
either \fexxv\ or \fexxvi\ \ka\ (6.67 and 6.97\,\kev, respectively) the most likely. This
would imply an outflow velocity of $\sim$0.2$c$ for \fexxvi, or an even more extreme
velocity of $\sim$0.25$c$ for \fexxv. No other features provide such a strong
improvement in the fit.

\begin{figure}
\hspace*{-0.35cm}
\epsscale{1.16}
\plotone{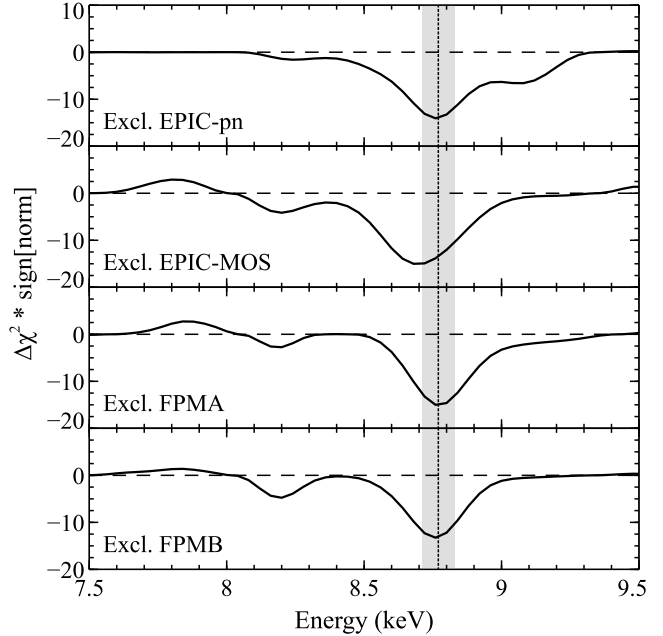}
\caption{
The results obtained repeating our line-search analysis after excluding each of the
detectors utilized in turn. In all cases, the same feature is picked out. The significance
is lower, as expected given the loss of S/N, but the improvement is still always
$\Delta\chi^{2}=13-15$ (for 3 extra DoF). The dotted line and grey shaded region
indicate the best-fit line energy and its uncertainty from the full analysis incorporating
all detectors.
}
\vspace{0.2cm}
\label{fig_excl}
\end{figure}

\begin{figure*}
\hspace*{-0.5cm}
\epsscale{1.16}
\plotone{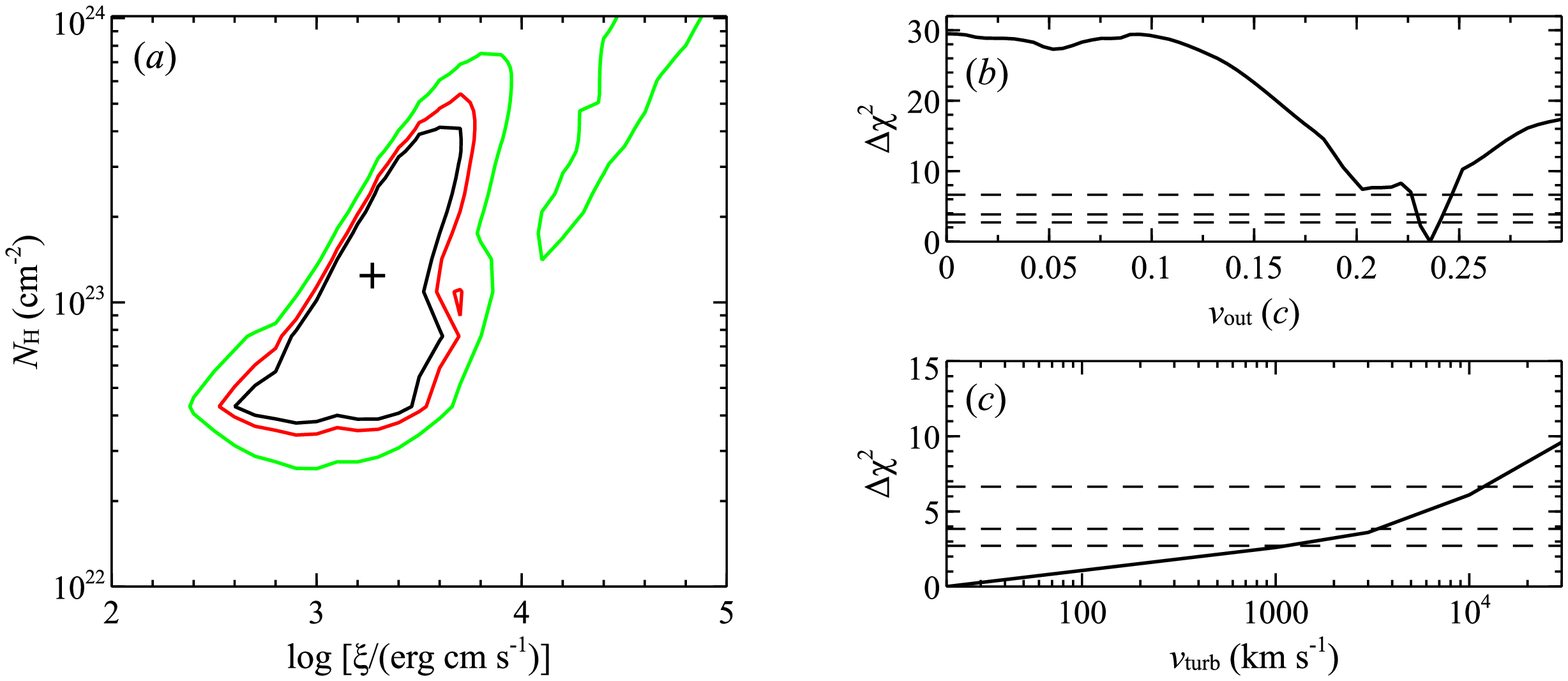}
\caption{
Confidence contours for the column density and ionisation parameter (2D;
\textit{panel a}), the outflow velocity (\textit{panel b}), and the turbulent velocity
broadening (\textit{panel c}) obtained from our photoionisation modeling with
\xstar. Panel a shows the 90, 95 and 99\% contours for two parameters of interest,
and the dashed lines in panels b and c show these same confidence levels for a
single parameter.
}
\vspace{0.2cm}
\label{fig_xstar}
\end{figure*}

In Figure \ref{fig_ratio} we show the data/model ratios for the model excluding and
including this absorption line. The feature is not particularly visually prominent, and
would be lost in the noise for any of the detectors individually. However, the key to
the statistical improvement observed is that all of the detectors show low residuals
to the continuum model at the same energy. In Figure \ref{fig_excl} we show the
same line search applied excluding each of the four detectors utilized in this work
in turn. In each of these cases, the improvement provided by including an
absorption line at $\sim$8.8\,keV is still $\Delta\chi^{2}=13-15$. This strongly
implies that this feature cannot be related to systematic effects (\eg\ instrumental
background), which differ for each of these detectors.

\subsection{Significance Simulations}

In order to assess the statistical significance of this potential Fe K absorption
feature, we performed a series of spectral simulations. Using the same response
and  background files, and adopting the same exposure times as the real data
used here, we simulated \nsims\ sets of \xmm\ (pn, combined MOS1 and MOS2)
and \nustar\ (FPMA, FPMB) spectra with the {\small FAKEIT} command in XSPEC
based on the best-fit cutoff powerlaw continuum (\ie\ without any absorption
feature included). Each of the simulated datasets was rebinned in the same
manner and analysed over the same bandpass as adopted for the real data. We
then fit each of the combined datasets with a cutoff powerlaw continuum, and
subsequently applied an identical line search as performed above (thus the
number of energy bins searched is fully accounted for). Of the \nsims\ datasets
simulated, only \ngtr\ returned a chance improvement equivalent to or greater
than that observed (at any energy searched), implying that the feature seen in
the real data is significant at the $\sim$3$\sigma$ level. This is something of a
conservative estimate; if we also require that the simulations match the exclusion
trial results given above (such that the deviation is not dominated by a single
detector), only \ngtrexcl\ match the observed characteristics by chance, indicating
a $\sim$99.9\% detection significance. Using the velocity information from
\cite{Pinto16nat} as a prior would also serve to further increase the detection
significance.

\subsection{XSTAR Modeling}

We model this absorption feature using a physical model for absorption by
a photoionised plasma. Following \cite{Middleton14}, we construct a set of custom
photoionized absorption models with \xstar\ (\citealt{xstar}), using the continuum
emission observed from \ngc\ (described above) as the input ionizing continuum.
These grids are calculated assuming solar abundances, an ionizing luminosity of
10$^{40}$\,\ergps\ (typical for \ngc), and a density of $10^{17}$\,cm$^{-3}$ (see
\citealt{Middleton14}). Free parameters are the ionisation parameter
($\xi=L_{\rm{ion}}/nR^{2}$, where $L_{\rm{ion}}$ is the ionizing luminosity between
1--1000\,Ry, $n$ is the density of the plasma and $R$ is its distance from the
ionizing source; $\xi$ is calculated in units of \ergcmps\ throughout this work), the
column density ($N_{\rm{H,ion}}$) and the line-of-sight outflow velocity
($v_{\rm{out}}$) of the absorbing medium. We also consider a range of turbulent
velocity broadening, $v_{\rm{turb}}$, from 20--30,000\,\kmps, for consistency with
\cite{Pinto16nat}.

The addition of \xstar\ to our continuum model provides a more substantial
improvement to the fit than the single Gaussian feature: $\Delta\chi^{2}=29$ for three
extra free parameters. This may suggest that, in addition to the feature at
$\sim$8.8\,keV, there are further weak features in the spectrum that are associated
with the same outflow but not significantly detected individually. We find
$\log\xi=3.3^{+0.3}_{-0.5}$,
$N_{\rm{H,ion}}=(1.2^{+2.1}_{-0.8})\times10^{23}$\,cm$^{-2}$,
$v_{\rm{out}}=0.236\pm0.005c$, and $v_{\rm{turb}}<1000$\,\kmps. Although
$\log\xi\sim3.3$ is preferred, there is a degeneracy between the ionisation
parameter and the column density (Figure \ref{fig_xstar}). This is not surprising
given that the observational signature of this absorber is dominated by a single line,
making it difficult to distinguish between ionisation states dominated by \fexxv\
($\log\xi\sim3.3$) and \fexxvi\ ($\log\xi\sim4.5$; see \eg\ \citealt{King14}). The
velocity contour therefore also shows two minima with similarly good fits (the second
being at $v_{\rm{out}}\sim0.2c$; Figure \ref{fig_xstar}), corresponding to these two
potential solutions.

\section{Discussion and Conclusions}
\label{sec_dis}

We have presented the detection of an absorption feature at $E=8.77^{+0.05}_{-0.06}$
\kev\ in the X-ray spectrum of the ULX \ngc, found by combining data from the \xmm\
and \nustar\ observatories. Owing to the extreme energy of this feature, and the low
flux of \ngc, the combination of \xmm\ and \nustar\ is particularly vital to this detection.
This provides a broad bandpass, enabling robust continuum estimation both above
and below the line energy, and significantly enhances the S/N over what each
observatory individually would return in commensurate exposure times. Both of
these issues hindered our previous attempt to search for absorption in \ngc\ (using
\xmm\ only; \citealt{Walton12ulxFeK}) to the extent that this feature could not be
seen. Furthermore, the combination of the different detectors aboard \xmm\ and
\nustar\ allows us to effectively rule out an instrumental systematic origin, given that
all the detectors utilized show consistent low residuals to the continuum emission.

Associating this feature with highly ionised iron, either \fexxv\ or \fexxvi, implies an
extreme outflow velocity of 0.2--0.25$c$. Photoionisation modeling marginally
prefers a solution in which this absorption is dominated by \fexxv\
($\log\xi\sim3.3$, $v_{\rm{out}}\sim0.25c$), but with the observational signature of
this absorber being dominated by this single line there is significant degeneracy, 
with solutions dominated by \fexxvi\ ($\log\xi\sim4.5$, $v_{\rm{out}}\sim0.2c$)
providing similarly good fits. These velocities are consistent with the UFO recently 
discovered by \cite{Pinto16nat}, that was identified through the
detection of highly blue-shifted absorption lines from moderately ionised material
in the low-energy X-ray band, suggesting that we are seeing an iron K$\alpha$
component associated with the same outflow. Ionized iron K$\alpha$ absorption
features associated with UFOs ($v_{\rm{out}}>0.1c$) have been seen in
several active galaxies (\eg\ \citealt{Tombesi10b}), but never before from an X-ray
binary. We note that the energy of the detected feature is just about consistent with
the rest-frame energy of the \fexxv\ edge at 8.83\,keV. However, the data prefer the
feature to be narrow, and no corresponding absorption line is seen at 6.67\,keV,
hence an ionized absorber at rest provides a significantly worse fit in our
photoionization modeling ($\Delta\chi^{2}\sim29$).

\cite{Pinto16nat} consider two possibilities for the outflow structure: a single zone
with low velocity broadening ($v_{\rm{turb}}=20$\,\kmps), and two zones, the
second of which has a much higher broadening ($v_{\rm{turb}}=10,000$\,\kmps).
The line detected here is narrow; broadening at the latter level seems to be unlikely.
We therefore compare our results to the former scenario. Although there is
significant degeneracy in our results (Figure \ref{fig_xstar}), the absorption detected
here is significantly more ionized, and has a significantly larger column;
\citealt{Pinto16nat} found $\log\xi\sim2.3$, and
$N_{\rm{H}}\sim2\times10^{22}$\,cm$^{-2}$ for their one-zone model. The
absorption detected here may thus arise in a phase of the outflow located closer to
the black hole than that contributing the features detected in the RGS. The contrast
between $v_{\rm{out}}$ and $v_{\rm{turb}}$ is larger than inferred for the UFOs in
PDS456 (\citealt{Nardini15}) and PG1211+143 (\citealt{Pounds03}), despite the
similar Gaussian line width constraints. If real, this may provide some clue to the
wind geometry, implying that we might not directly view the primary acceleration
region, otherwise a smaller contrast would have been expected. However, the
constraint on $v_{\rm{turb}}$ is ionization dependent, with the higher ionization
solution allowing for $v_{\rm{turb}}$ up to 10,000\,\kmps, more comparable with
these other cases.

This additional phase of absorption would significantly increase the total mass
outflow rate ($\dot{M}_{\rm{out}}$) compared to that inferred from just the low-energy
absorption alone. Combining the standard expression for $\dot{M}_{\rm{out}}$ and
the definition of the ionisation parameter, we can estimate the kinetic luminosity of the
outflow ($L_{\rm{kin}}=1/2\dot{M}v_{\rm{out}}^2$) relative to the bolometric radiative
luminosity ($L_{\rm{bol}}$):

\begin{equation}
\frac{L_{\rm{kin}}}{L_{\rm{bol}}}\approx2{\pi}m_{\rm{p}}\mu\frac{L_{\rm{ion}}}{L_{\rm{bol}}}\frac{v_{\rm{out}}^3}{\xi}{\Omega}C_{\rm{V}}
\end{equation}

\noindent{where} $m_{\rm{p}}$ is the proton mass, $\mu$ is the mean atomic
weight ($\sim$1.2 for solar abundances), $\Omega$ is the (normalized) solid angle
subtended by the wind, and $C_{\rm{V}}$ is its volume filling factor (or its
`clumpiness'). Although some extrapolation beyond the observed bandpass is
necessary, the broadband continuum models constructed by \cite{Bachetti13} and
\cite{Miller14} imply $L_{\rm{ion}}/L_{\rm{bol}}\sim0.85$. We therefore find
$L_{\rm{kin}}/L_{\rm{bol}}\sim1500{\Omega}C_{\rm{V}}$ and
$\sim$60${\Omega}C_{\rm{V}}$ for the lower and higher ionisation solutions,
respectively. Thus unless it either has a very small solid angle or a very low
volume filling factor (which may be possible; \citealt{King12}), the wind may dominate
the energy output from \ngc. While both $\Omega$ and $C_{\rm{V}}$ are unknown,
the above $L_{\rm{kin}}/L_{\rm{bol}}$ values are extreme in comparison to similar 
calculations for even the strongest outflows seen from sub-Eddington systems
(\citealt{Blustin05, King12, King14, Nardini15, Miller16}). This is consistent with the
basic expectation for a super-Eddington accretion scenario (\citealt{Poutanen07,
King09}), as suggested by the unusual broadband X-ray spectrum observed
(\citealt{Bachetti13}).

\section*{ACKNOWLEDGEMENTS}

The authors would like to thank the anonymous referee for their extremely timely
and positive feedback, which helped improve the final manuscript. MJM
acknowledges support from an STFC Ernest Rutherford fellowship, CP and ACF
acknowledge support from ERC Advanced Grant 340442, and DB acknowledges
financial support from the French Space Agency (CNES). This research has made
use of data obtained with \nustar, a project led by Caltech, funded by NASA and
managed by NASA/JPL, and has utilized the \nustardas\ software  package, jointly
developed by the ASDC (Italy) and Caltech (USA). This research has also made
use of data obtained with \xmm, an ESA science mission with instruments and 
contributions directly funded by ESA Member States.

{\it Facilites:} \facility{NuSTAR}, \facility{XMM}

\bibliographystyle{/Users/dwalton/papers/mnras}

\bibliography{/Users/dwalton/papers/references}

\label{lastpage}

\end{document}